%% file: ampls.tex
\documentstyle[epsf,eqsecnum,prc,preprint,aps]{revtex}
\begin{document}
\tighten
\draft
\renewcommand{\thefootnote}{\fnsymbol{footnote}}
\title
{\bf Chiral Lagrangians and the transition amplitude for radiative muon capture
}
\author{J.\,Smejkal and E.\,Truhl\'{\i}k}
\address{
Institute of Nuclear Physics, Czech Academy of Sciences, CZ-25068
$\check{R}$e$\check{z}$, Czech Republic\\
\vspace{-15pt}
\begin{center}
and
\end{center}
\vspace{-30pt}
}
\author{F.\,C.\,Khanna
}
\address{Theoretical Physics Institute, Department of Physics, University
of Alberta, Edmonton, Alberta,Canada,T6G 2J1\\
\vspace{-15pt}
\begin{center}
and
\end{center}
\vspace{-15pt}
TRIUMF, 4004 Wesbrook Mall, Vancouver, BC, Canada, V6T 2A3\\}
\maketitle
\begin{abstract}
\input{abs}
\end{abstract}
\input feynman
\input{intro}
\input{chapter1}
\input{chapter2}

\input{chapter3}

\acknowledgments
\input{ack}
\input{ref}

\input{fig}
\end{document}

%% file: abs.tex

The transition operator for the radiative capture of mesons $\mu^-$ by protons
is constructed starting from a chiral Lagrangian of the $N \pi \rho\,a_1\,
\omega$ system obtained within the approach of hidden
local symmetries. The transition operator is gauge invariant and 
satisfies exactly the CVC and PCAC equations.

%% file: intro.tex
\newpage

\section{Introduction}
\label{intro}

The radiative muon capture (RMC) on proton,
\begin{equation}
\mu^-\,+\,p\,\longrightarrow\,n\,+\,\nu_\mu\,+\,\gamma\,,  \label{mupnnug}
\end{equation}
has recently been measured at TRIUMF \cite{TRIUMF} for the first time.
The aim of the experiment was to extract the value of the weak induced
pseudoscalar constant $g_P$ with the result
\begin{equation}
g_P(q^2\,\approx\,0.88\,m^2_\mu)\,=\,(9.8\,\pm\,0.7\,\pm\,0.3)\,g_A(0)\,.
\label{gPexp}
\end{equation}
This value of $g_P$ is about 1.5 times larger than tho one predicted by PCAC
and admitting pion-pole dominance of the induced pseudoscalar part of the weak
axial current. Actually, the constant $g_P$ is the only one not well
known experimentally of the four constants $g_V$,$g_M$,$g_A$ and $g_P$
entering the weak nucleon current.

There are constant efforts for many years to extract $g_P$ from the ordinay
muon capture by proton,
\begin{equation}
\mu^-\,+\,p\,\longrightarrow\,n+\nu_\mu\,,  \label{OMCP}
\end{equation}
with the world average value \cite{BAR} charged with a 25\% error. The recent
precise measurement \cite{PSI} of the transition rate for the reaction
\begin{equation}
\mu^-\,+\,^{3}He\,\longrightarrow\,^{3}H\,+\,\nu_\mu\,,
\end{equation}
lead subsequently \cite{CF,CT} to the extraction of $g_P$ with an accuracy
of $\approx$ 19\%,
\begin{equation}
g_P\,=\,(1.05\,\pm\,0.19)\,g^{PCAC}_P\,,
\end{equation}
with
\begin{equation}
g^{PCAC}_P(q^2)\,=\,\frac{2Mm_\mu}{q^2+m^2_\pi}\,g_A(q^2)\,.
\end{equation}
For reaction (\ref{OMCP}), $q^2\,\approx\,0.88\,m^2_\mu$ and
$g^{PCAC}_P\,\approx\,6.6\,g_A$. On the other side, the value of $q^2$
in the hadron radiative part of the amplitude for RMC can reach
$q^2\,\approx\,-m^2_\mu$ at the high end of the photon spectrum,
which leads to an enhancement by a factor of $3$ in
the amplitude due to the induced pseudoscalar part of the weak axial
interaction
in this kinematical region.
It is this feature of the RMC process which feeds the hope that it can be
effectively used for extracting of the value of $g_P$.

The transition amplitude for the elementary reaction (\ref{mupnnug})
was derived by numbers of authors. One can use \cite{RT} Feynman graphs obtained
by attaching a $\gamma$ in every  possible way to the lines of
electromagnetically interacting particles participating in reaction
(\ref{mupnnug}). Then the current conservation allows one to get the
hadron radiative amplitude up to terms linear in photon momentum $k$.
Better way to get the amplitude is using the low energy theorems
(see Refs.\,\cite{AD}-\cite{BF} and references therein) which provide
the RMC amplitude in terms of elastic
weak form factors and pion photoproduction amplitudes up to terms linear in
$k$ and $q=p_\mu\,-\,p_\nu$. The gauge invariance, CVC and PCAC are
respected in this approach just to this order. Such an amplitude
was applied in \cite{TRIUMF} to extract the constant $g_P$
of Eq.\,(\ref{gPexp}).

The difference between the value (\ref{gPexp}) of $g_P$ and its PCAC
prediction opens naturally among several questions also
the discussion about the structure and
completeness of the applied transition amplitude. Here we present a
new one obtained from the Lagrangian of the $N \pi \rho\,a_1\,\omega$
system constructed within the approach of hidden local symmetries (HLS)
\cite{STG}. The advantage of this amplitude is that it satisfies gauge
invariance, CVC and PCAC exactly.

In the approach of HLS \cite{BKY,M}, a given global symmetry group $G_g$
of a system Lagrangian is extended to a larger one by a local group
$H_l$ and the Higgs mechanism generates the mass terms for gauge fields
of the local group in such a way that the local symmetry is conserved.
For the chiral group $G_g\,\equiv\,[SU(2)_L \times SU(2)_R]_{g}$ and
$H_l\, \equiv\, [SU(2)_L \times SU(2)_R]_{l}$ the gauge particles are
identified \cite{BKY,M} with the $\rho$- and $a_1$ mesons.
An additional extension by the group $U(1)_l$ allows one to include
the isoscalar $\omega$ meson as well \cite{KM}. Moreover, external
gauge fields, which are related  to the electroweak interactions
of the Standard Model, are included by gauging the
global chiral symmetry group $G_g$.

In Sect.\,\ref{CH1}, we write down the HLS Lagrangian and
the associated currents necessary to construct the transition
amplitude for RMC.
In Sect.\,\ref{CH2}, we construct this amplitude and show that
it is gauge invariant and that it satisfies CVC and
PCAC exactly.
In Sect.\,\ref{CH3}, we compare our amplitude with the one
obtained from the low energy theorems and we make our conclusion.

%% file: chapter1.tex

\section{Lagrangian and the currents \label{CH1}}

We use the Lagrangian written in terms of the heavy meson fields 
representing the nonlinear realization of the HLS for the groups
$H_l\,\equiv\, [SU(2)_L \times SU(2)_R]_{l} \times U(1)_l$
 \cite{STG,KM}. It can be written as
\begin{eqnarray}
{\cal L}_{N \pi \rho\,a_1\,\omega} & = & -{\bar N}\gamma_\mu
\partial_\mu N -M{\bar N}N+ig{\bar N} \gamma_5
(\vec{\Pi}\,\cdot\,\vec{\tau})N
- ig_\rho \frac{g_A}{2f_\pi}{\bar N}\gamma_\mu \gamma_5
(\vec{\tau}\,\cdot\,\vec{\rho}_\mu \times \vec{\Pi})N \nonumber  \\
& & -i\frac{g_\rho g^2_A}{f_\pi} {\bar N} \gamma_\mu(\vec{\tau}\,\cdot\,
 {\vec a}_\mu \times \vec{\Pi})N
-ig_A g_\rho {\bar N}\gamma_\mu \gamma_5 (\vec{\tau}\,\cdot\,
\vec a_\mu)N  \nonumber \\
& & -i\frac{g_\rho}{2}{\bar N}(\gamma_\mu \vec{\rho}_\mu
-i\frac{\kappa_V}{4M}\sigma_{\mu \nu}\vec{\cal F}^{(\rho)}_{\mu \nu})
\,\cdot\,\vec{\tau}N 
 -i\frac{g_\omega}{2}{\bar N}(\gamma_\mu \omega_\mu
-i\frac{\kappa_S}{4M}\sigma_{\mu \nu} \omega_{\mu \nu})N 
\nonumber \\
& & +i\frac{g_\rho g_A}{4f_\pi}\frac{\kappa_V}{2M}\,{\bar
N}\gamma_5 \sigma_{\mu \nu} (\vec{\Pi}\cdot \vec{\cal F}^{(
\rho)}_{\mu \nu})N
+i\frac{g_\omega g_A}{4f_\pi}\frac{\kappa_S}{2M}\,{\bar
N}\gamma_5 \sigma_{\mu \nu} (\vec{\Pi}\cdot \vec{\tau})\omega_{\mu \nu}N
\nonumber \\
& & + g_\rho \vec{\rho}_\mu \cdot \vec{\Pi} \times \partial_\mu \vec{\Pi} 
-g_\rho \partial_\nu \vec{\rho}_\mu \cdot \vec{\rho}_\mu \times 
\vec{\rho}_\nu + g_\rho (\vec{\rho}_\mu \times \vec{a}_\nu -
\vec{\rho}_\nu \times \vec{a}_\mu) \cdot \partial_\mu \vec{a}_\nu
\nonumber \\
& & + \frac{1}{f_\pi} (\vec{\rho}_{\mu \nu} \cdot \vec{a}_\mu \times 
\partial_\nu \vec \Pi + \frac{1}{2} \vec{\rho}_\mu \cdot
\partial_\nu \vec{\Pi} \times \vec{a}_{\mu \nu}) \nonumber \\
& & + {\cal O}(|\vec{\Pi}|^2)\,. \label{LNLNPNPR}
\end{eqnarray}
Here
\begin{equation}
\vec{\cal F}^{(\rho)}_{\mu \nu} = \vec{\rho}_{\mu \nu} - g_\rho
\vec{\rho}_\mu \times \vec{\rho}_\nu\,,\qquad \vec{\rho}_{\mu \nu} = 
\partial_\mu \vec{\rho}_\nu - \partial_\nu \vec{\rho}_\mu\,,\qquad
\omega_{\mu \nu} = \partial_\mu \omega_\nu - \partial_\nu \omega_\mu\,.
\end{equation}
The Lagrangian (\ref{LNLNPNPR}) describes reasonably all the relevant 
elementary processes ($\rho \pi \pi$, $a_1\,\gamma \pi$ etc.\,) at
intermediate energies.

The associated currents are obtained by the Glashow--Gell-Mann method 
\cite{GGM} and read
\begin{eqnarray}
J^S_{V\,,\,\mu} & = & -\frac{m^2_\omega}{g_\omega}\, \omega_\mu\,,
\label{JVS} \\
\vec{J}_{V\,,\,\mu} & = & -\frac{m^2_\rho}{g_\rho}\, \vec{
\rho}_\mu-2f_\pi g_\rho \, {\vec a}_\mu
\times \vec{\Pi}+{\cal O}(|\vec{\Pi}|^2)\,,  \label{JVV}  \\
\vec{J}_{A\,,\,\mu} & = &
-\frac{m^2_\rho}{g_\rho}\,\vec{a}_\mu
+f_\pi \partial_\mu \vec{\Pi}-2f_\pi g_\rho\,
\vec{\rho}_\mu \times \vec{\Pi} \nonumber \\
& & +\frac{1}{g_\rho}\,(\frac{1}{2f_\pi}\, \partial_\nu \vec{\Pi}
-g_\rho \vec{a}_\nu + e\vec{\cal A}_\nu) \times
\vec{\rho}_{\mu \nu} + {\cal O}(|\vec{\Pi}|^2)\,. \label{JA}  
\end{eqnarray}
We now have the full set of vertices and currents which are necessary to
construct the transition amplitude for RMC at the tree level.

%% file: chapter2.tex

\section{Transition amplitude for RMC \label{CH2}}

The needed transition operator $T^1-iT^2$ consists of the following terms
\begin{equation}
T^a(k,q)  =  \frac{eG}{\sqrt{2}}\,\{\,M^a(k,q)\,+\,l_\mu(0)\epsilon^\star_\nu(k)
\,[\,M^{B,\,a}_{\mu \nu}(k,q)\,+\,M^a_{\mu \nu}(\pi;k,q)\,+\,
M^{c.\,t.,\,a}_{\mu \nu}(a_1;k,q)\,]\,\}\,, \label{Ta}
\end{equation}
where $M^a(k,q)$ corresponds to the radiation by muon and the radiative hadron
amplitude is given in the square braces as a sum of three terms. As it will
become clear later, these amplitudes present actually a set of terms which
satisfy separately a closed continuity equation which together
provide the PCAC for the radiative hadron amplitude. In particular,
$M^{B,\,a}_{\mu \nu}$ contains
the nucleon Born terms (Figs.\,1a and 1b) and some related
contact amplitudes (Figs.\,1c and 1d),
$M^a_{\mu \nu}(\pi)$ contains the mesonic amplitude
$M^{m.\,c.\,,a}_{\mu \nu}(k,q)$ describing the
radiation of the photon by the pion in flight which was created by the weak
axial current and all contact terms where an electroweak vertex
(the bubble on the graph) is
connected to the nucleon via pion line (Fig.\,1e with the pion).
Similarly, $M^{c.\,t.,\,a}_{\mu \nu}(a_1)$
is the sum of all contact terms where the electroweak  interaction is
connected to the nucleon by $a_1$ meson line (Fig.\,1e with the
$a_1$ meson).

We construct the transition amplitude for RMC  using the Feynman graphs.
Generally, any of amplitudes $M^a_{\mu \nu}(k,q)$ given below is
related to the corresponding S-matrix element as
\begin{equation}
S\,=\,(2\pi)^4\,i\,\delta^4(k+q_1-q)\,l_\mu(0)\epsilon^\star_\nu(k)\,
     M^a_{\mu \nu}(k,q)\,.
\end{equation}

The radiation by muon is
\begin{eqnarray}
M^a(k,q) & = & -i\bar u (p_\nu)\,\gamma_\mu(1+\gamma_5)
S_F(p_{\mu^-} - k) i \epsilon^*_\nu(k) \gamma_\nu\,u(p_{\mu^-})\,
\nonumber \\
& & \bar u(p')\,\hat J^a_{\,W,\,\mu}(q_1)\,u(p)\,, \label{Ma} \\
\hat J^a_{\,W,\,\mu}(q_1) & = & \hat J^a_{\,V,\,\mu}(q_1) +
\hat J^a_{\,A,\,\mu}(q_1)\,, \label{JaW} \\
\hat J^a_{\,V,\,\mu}(q_1) & = & i m^2_\rho \Delta^\rho_{\mu \eta}
(q_1)(\gamma_\eta - \frac{\kappa_V}{2M} \sigma_{\eta \delta}
q_{1 \delta})\,\frac{\tau^a}{2}\,, \label{JaV} \\
\hat J^a_{\,A,\,\mu}(q_1) & = & i[-g_A m^2_{a_1}\Delta^{a_1}_{\mu
\nu}(q_1)\gamma_\nu \gamma_5\,+\,2iMg_A\,\Delta^\pi_F(q_1)
q_{1 \mu} \gamma_5]\,\frac{\tau^a}{2}\,, \label{JaA} \\
\Delta^B_{\mu \nu}(l) & = & (\delta_{\mu \nu} + \frac{l_\mu
l_\nu}{m^2_B})\,\Delta^B_F(l)\,,\qquad B=\rho,\,a_1\,, \\
q_1 & = & p'-p\,,\qquad S_F(l) = -\frac{1}{i\,l_\mu \gamma_\mu+M}\,.
\end{eqnarray}
Here $g_A=-1.26$ and our definition of the electromagnetic and
weak currents conforms Ref.\,\cite{BS}.

Let us now present the radiative hadron amplitude. The part
$M^{B,\,a}_{\mu \nu}(k,q)$ is
\begin{eqnarray}
M^{B,\,a}_{\mu \nu}(k,q)& = & \sum_{i=1}^6\,M^{B,\,a}_{\mu
 \nu}(i;k,q)\,, \label{MBa} \\
M^{B,\,a}_{\mu \nu}(1) & = & -{\bar u}(p')[\,\hat J^a_{W,\,\mu}(q)
S_F(Q)\hat J^{e.\,m.}_\nu(k)\,+\,\hat J^{e.\,m.}_\nu(k) S_F(P)
\hat J^a_{W,\,\mu}(q)\,]u(p)  \nonumber \\
& & \equiv\,M^{B,a}_{V,\,\mu \nu}(1)\,+\,M^{B,a}_{A,\,\mu \nu}(1)\,,
\label{MBa1} \\
M^{B,\,a}_{\mu \nu}(2) & = & -\frac{g_A}{2} m^2_\rho
\Delta^\rho_{\nu \zeta}(k) q_\mu \Delta^\pi_F(q)
\varepsilon^{3\,a\,b} \Gamma^b_{\zeta}(p',p) \nonumber \\
&\equiv & \,
-i f_\pi q_\mu \Delta^\pi_F(q)\,{\cal M}^{B,a}_{\pi,\,\nu}(2)\,,
\label{MBa2} \\
M^{B,\,a}_{\mu \nu}(3) & = & i\frac{g_A}{2} \frac{\kappa_V}{2M}
m^2_\rho \Delta^\rho_{\nu \eta}(k) q_\mu \Delta^\pi_F(q) k_\zeta
\delta_{3\,a} \bar u(p')\gamma_5\,\sigma_{\zeta \eta} u(p)
\nonumber \\
& \equiv & \,
-i f_\pi q_\mu \Delta^\pi_F(q)\,{\cal M}^{B,a}_{\pi,\,\nu}(3)\,,
\label{MBa3} \\
M^{B,\,a}_{\mu \nu}(4) & = & -i\frac{g_A}{2} \frac{\kappa_S}{2M}
m^2_\omega \Delta^\omega_{\nu \eta}(k) q_\mu \Delta^\pi_F(q) k_\zeta
\bar u(p')\gamma_5\,\sigma_{\zeta \eta}\tau^a u(p) \nonumber \\
& \equiv & \,
-i f_\pi q_\mu \Delta^\pi_F(q)\,{\cal M}^{B,a}_{\pi,\,\nu}(4)\,,
\label{MBa4} \\
M^{B,\,a}_{\mu \nu}(5) & = & \frac{m^4_\rho}{2}\frac{\kappa_V}{2M}
\Delta^\rho_{\mu \eta}(q)\Delta^\rho_{\nu \zeta}(k)
\varepsilon^{3\,a\,b}
\bar u(p')\sigma_{\zeta \eta} \tau^b u(p)\,, \label{MBa5} \\
M^{B,\,a}_{\mu \nu}(6) & = & -i m^2_\rho \varepsilon^{3\,a\,b}\,
\bar u(p')[\,(q_\zeta+k_\zeta)\Delta^\rho_{\mu \eta}(q)
\Delta^\rho_{\nu \eta}(k)\,\hat J^b_{V,\,\zeta}(q_1) \nonumber \\
& & -\,(q_{1 \zeta}+q_\zeta)\Delta^\rho_{\mu \eta}(q)
\Delta^\rho_{\nu \zeta}(k)\,\hat J^b_{V,\,\eta}(q_1)\,+\,
(q_{1 \zeta}-k_\zeta)\Delta^\rho_{\mu \zeta}(q)
\Delta^\rho_{\nu \eta}(k)\,\hat J^b_{V,\,\eta}(q_1)\,]u(p)\,.
\label{MBa6}
\end{eqnarray}
We write also down the contribution from the radiative nucleon Born
term due to the induced pseudoscalar
\begin{eqnarray}
M^{B,\,a}_{ps,\,\mu \nu}(k,q) & = & Mg_A q_\mu \Delta^\pi_F(q)\,
{\bar u}(p')[\,\gamma_5 \tau^a
S_F(Q)\hat J^{e.\,m.}_\nu(k)\, \nonumber \\
& & +\,\hat J^{e.\,m.}_\nu(k) S_F(P)
\gamma_5 \tau^a\,]u(p)\,\equiv\,-i f_\pi q_\mu \Delta^\pi_F(q)\,{\cal
M}^{B,a}_{\pi,\,\nu}(1)\,,\label{MBpsa}
\end{eqnarray}
which serves to define the radiative pion absorption amplitude
${\cal M}^{B,a}_{\pi,\,\nu}(1)$. Besides presenting contact
amplitudes, eqs.\,(\ref{MBa2}),(\ref{MBa3})
and (\ref{MBa4}) define also the radiative pion absorption
amplitudes ${\cal M}^{B,a}_{\pi,\,\nu}(2)$,
${\cal M}^{B,a}_{\pi,\,\nu}(3)$ and ${\cal
M}^{B,a}_{\pi,\,\nu}(4)$, respectively. In Eq.\,(\ref{MBa1}),
the amplitudes $M^{B,a}_{V,\,\mu \nu}(1)$ and $M^{B,a}_{A,\,\mu
\nu}(1)$ correspond to the vector-vector and axial-vector part of the
weak nucleon current (\ref{JaW}).

In Eqs.\,(\ref{MBa1}-\ref{MBpsa}) the following notations are
used
\begin{eqnarray}
\hat J^{e.\,m.}_\nu(k)\,& = &\,\hat J^S_{V,\,\nu}(k)\,+\,\hat
J^3_{V,\,\nu}(k)\,,  \label{Jem} \\
\hat J^S_{V,\,\nu}(k)\,& = &\,i m^2_\omega \Delta^\omega_{\nu
\eta}(k)\frac{1}{2}\,(\gamma_\eta\,-\,\frac{\kappa_S}{2M}\,
\sigma_{\lambda \eta} k_\lambda)\,, \label{JS} \\
\hat J^3_{V,\,\nu}(k)\,& = &\,i m^2_\rho \Delta^\rho_{\nu
\eta}(k)\frac{\tau^3}{2}\,(\gamma_\eta\,-\,\frac{\kappa_V}{2M}\,
\sigma_{\lambda \eta} k_\lambda)\,, \label{JemV} \\
\Gamma^b_\nu(p',p)\, & = & \,\bar u(p')\,\gamma_5 \gamma_\nu
\tau^b\,u(p)\,. \label{Gbnu}
\end{eqnarray}

We keep the amplitudes (\ref{MBa1})-(\ref{MBpsa}) together
because they satisfy the following continuity equation
\begin{equation}
q_\mu\,M^{B,a}_{\mu \nu} = if_\pi m^2_\pi \Delta^\pi_F(q)\,
\sum_{i=1}^4\,{\cal M}^{B,\,a}_{\pi,\,\nu}(i)\,+\,
i\varepsilon^{3\,a\,b}\bar u(p')\,\hat J^b_{V,\,\mu}(q_1)\,
u(p)\,. \label{PCAC1}
\end{equation}

We further present  the second part of the radiative hadron amplitude
$M^a_{\mu \nu}(\pi;k,q)$ as
\begin{equation}
M^a_{\mu \nu}(\pi;k,q)\, = \,M^{m.\,c.\,,a}_{\mu \nu}(k,q)\,+\,
\sum_{i=1}^5\,M^a_{\mu \nu}(\pi,i;k,q)\,. \label{Mapi}
\end{equation}
Here
\begin{eqnarray}
M^{m.\,c.\,,a}_{\mu \nu} & = & -Mg_A q_\mu \Delta^\pi_F(q) m^2_\rho
\Delta^\rho_{\eta \nu}(k)\,(q_{1 \eta}+q_\eta)
\Delta^\pi_F(q_1)\,\Gamma^a(p',p) \nonumber  \\
& \equiv & \,-if_\pi q_\mu \Delta^\pi_F(q)
{\cal M}^{m.\,c.\,,a}_{\pi,\,\nu}\,, \label{Mmca} \\
M^a_{\mu \nu}(\pi,1) & = & -2i Mg_A m^2_\rho \Delta^{a_1}_{\mu
\nu}(q)\,\Delta^\pi_F(q_1)\varepsilon^{3\,a\,b}\Gamma^b(p',p)\,,
\label{Mapi1} \\
M^a_{\mu \nu}(\pi,2) & = & 2i Mg_A m^2_\rho \Delta^{a_1}_{\mu
\eta}(q)\Delta^\rho_F(k)(\,k_\eta q_{1 \nu} - q_1 \cdot k\,
\delta_{\eta \nu}\,)\,\Delta^\pi_F(q_1)\varepsilon^{3\,a\,b}\Gamma^b(p',p)\,,
\label{Mapi2} \\
M^a_{\mu \nu}(\pi,3) & = & i Mg_A m^2_\rho \Delta^{a_1}_F(q)
\Delta^\rho_{\eta \nu}(k)(\,q_\eta q_{1 \mu} - q \cdot q_1\,
\delta_{\eta \mu}\,)\,\Delta^\pi_F(q_1)\varepsilon^{3\,a\,b}\Gamma^b(p',p)\,,
\label{Mapi3} \\
M^a_{\mu \nu}(\pi,4) & = & 2i Mg_A m^2_\rho \Delta^\rho_{\mu
\nu}(k)\,\Delta^\pi_F(q_1)\varepsilon^{3\,a\,b}\Gamma^b(p',p)\,,
\label{Mapi4} \\
M^a_{\mu \nu}(\pi,5) & = & -i Mg_A \Delta^\rho_F(k)(\,k_\mu q_{1 \nu} -
q_1 \cdot k\,
\delta_{\mu \nu}\,)\,\Delta^\pi_F(q_1)\varepsilon^{3\,a\,b}\Gamma^b(p',p)\,,
\label{Mapi5} \\
\Gamma^b(p',p) & = & \bar u(p') \gamma_5 \tau^b u(p)\,.
\label{Gb}
\end{eqnarray}
The amplitude $M^a_{\mu \nu}(\pi;k,q)$ defined in
Eq.\,(\ref{Mapi}) satisfies the continuity equation
\begin{equation}
q_\mu M^a_{\mu \nu}(\pi;k,q) = i f_\pi m^2_\pi \Delta^\pi_F(q)
{\cal M}^{m.\,c.\,,a}_{\pi,\,\nu}
\,-\,i Mg_A q_{1 \nu} \Delta^\pi_F(q_1)
\varepsilon^{3\,a\,b}\Gamma^b(p',p)\,. \label{PCAC2}
\end{equation}
The radiative pion absorption amplitude
${\cal M}^{m.\,c.\,,a}_{\pi,\,\nu}$ is defined in
Eq.\,(\ref{Mmca}) and the second term at the r.\,h.\,s.\, of
Eq.\,(\ref{PCAC2}) is simply related to the induced pseudoscalar
part of the weak axial nucleon current  (\ref{JaA}).

The last amplitude of Eq.\,(\ref{Ta}) we need to discuss is
$M^{c.\,t.,\,a}_{\mu \nu}(a_1;k,q)$ which represents various
contact terms of the $a_1$ meson range (cf.\, Fig.\,1e with
$B=a_1$). Explicitly we have
\begin{eqnarray}
M^{c.\,t.,\,a}_{\mu \nu}(a_1;k,q) & = & \sum_{i=1}^5\,
M^{c.\,t.,\,a}_{\mu \nu}(a_1,i;k,q)\,, \label{Ma1} \\
M^{c.\,t.,\,a}_{\mu \nu}(a_1,1) & = & \frac{1}{2}g_A q_\mu
\Delta^\pi_F(q) m^2_{a_1}\Delta^{a_1}_{\nu \zeta}(q_1)
\varepsilon^{3\,a\,b} \Gamma^b_\zeta (p',p) \nonumber \\
& \equiv & \,
-if_\pi q_\mu \Delta^\pi_F(q)\,{\cal M}^{c.\,t.,\,a}_{\pi,\,\nu}
(a_1,1)\,, \label{Ma11} \\
M^{c.\,t.,\,a}_{\mu \nu}(a_1,2) & = & \frac{1}{2}g_A q_\mu
\Delta^\pi_F(q) m^2_{a_1}\Delta^{a_1}_{\eta \zeta}(q_1)
q_\lambda\,[\,k_\eta\Delta^\rho_{\lambda \nu}(k)-k_\lambda
\Delta^\rho_{\eta \nu}(k)\,]
\varepsilon^{3\,a\,b} \Gamma^b_\zeta (p',p) \nonumber \\
&\equiv & \,
-if_\pi q_\mu \Delta^\pi_F(q)\,{\cal M}^{c.\,t.,\,a}_{\pi,\,\nu}
(a_1,2)\,, \label{Ma12} \\
M^{c.\,t.,\,a}_{\mu \nu}(a_1,3) & = & -\frac{1}{2}g_A q_\mu
\Delta^\pi_F(q) m^2_\rho\Delta^\rho_{\eta \nu}(k)
q_\lambda\,[\,q_{1 \eta}\Delta^{a_1}_{\lambda \zeta}(q_1)-q_{1
\lambda}
\Delta^{a_1}_{\eta \zeta}(q_1)\,]
\varepsilon^{3\,a\,b} \Gamma^b_\zeta (p',p) \nonumber \\
&\equiv & \,
-if_\pi q_\mu \Delta^\pi_F(q)\,{\cal M}^{c.\,t.,\,a}_{\pi,\,\nu}
(a_1,3)\,, \label{Ma13} \\
M^{c.\,t.,\,a}_{\mu \nu}(a_1,4) & = & \frac{1}{2}g_A
m^2_{a_1}\Delta^{a_1}_{\lambda \zeta}(q_1)
[\,k_\mu\Delta^\rho_{\lambda \nu}(k)-k_\lambda
\Delta^\rho_{\mu \nu}(k)\,]
\varepsilon^{3\,a\,b} \Gamma^b_\zeta (p',p)\,, \label{Ma14} \\
M^{c.\,t.,\,a}_{\mu \nu}(a_1,5) & = & g_A m^4_\rho\Delta^\rho_{\eta \nu}(k)
\,[\,(q_\eta+q_{1 \eta})\Delta^{a_1}_{\lambda \mu}(q)
\Delta^{a_1}_{\lambda \zeta}(q_1)
-q_\lambda \Delta^{a_1}_{\eta \mu}(q) \Delta^{a_1}_{\lambda
\zeta}(q_1) \nonumber \\
& & + q_{1 \lambda} \Delta^{a_1}_{\lambda \mu}(q)
\Delta^{a_1}_{\zeta \eta}(q_1)\,]\,
\varepsilon^{3\,a\,b} \Gamma^b_\zeta (p',p)\,. \label{Ma15}
\end{eqnarray}

The continuity equation for the amplitude
$M^{c.\,t.,\,a}_{\mu \nu}(a_1;k,q)$
of Eq.\,(\ref{Ma1}) is
\begin{equation}
q_\mu M^{c.\,t.,\,a}_{\mu \nu}(a_1;k,q) = i f_\pi m^2_\pi \Delta^\pi_F(q)
\,\sum_{i=1}^3\,{\cal M}^{c.\,t.\,,a}_{\pi,\,\nu}(a_1,i)
\,+\,g_A m^2_\rho \Delta^{a_1}_{\zeta \nu}(q_1)
\varepsilon^{3\,a\,b}\Gamma^b_\zeta (p',p)\,. \label{PCAC3}
\end{equation}
The amplitudes ${\cal M}^{c.\,t.\,,a}_{\pi,\,\nu}(a_1,1-3)$ are
defined in Eqs.\,(\ref{Ma11})-(\ref{Ma13}) and the last term at
the r.\,h.\,s.\, of Eq.\,(\ref{PCAC3}) is simply related to the
contact part the weak axial nucleon current  (\ref{JaA}).

Summing up Eqs.\,(\ref{PCAC1}),(\ref{PCAC2}),(\ref{PCAC3})
provides the equation of the PCAC for the radiative hadron
amplitude
\begin{equation}
q_\mu\,[M^{B,\,a}_{\mu \nu}\,+\,M^a_{\mu \nu}(\pi)\,+\,
M^{c.\,t.,\,a}_{\mu \nu}(a_1)]\,=\,i f_\pi m^2_\pi
\Delta^\pi_F(q)\,{\cal M}^a_{\pi,\,\nu}\,+\,
i\varepsilon^{3\,a\,b} \bar u(p')\,\hat J^b_{W,\,\nu}(q_1)\,u(p)
\,, \label{PCAC}
\end{equation}
where the weak vector nucleon current $\hat J^b_{W,\,\mu}$
is defined in Eq.\,(\ref{JaW}) and
the full radiative pion absorption amplitude
${\cal M}^a_{\pi,\,\nu}$ is given by the sum of the partial
radiative pion absorption amplitudes discussed in connection
with Eqs.\,(\ref{PCAC1}),(\ref{PCAC2}),(\ref{PCAC3})
\begin{equation}
{\cal M}^a_{\pi,\,\nu}\,=\,\sum_{i=1}^4\,{\cal M}^{B,\,a}_{\pi,\,
\nu}(i)\,+\,{\cal M}^{m.\,c.,\,a}_{\pi,\,\nu}\,+\,
\sum_{i=1}^3\,{\cal M}^{c.\,t.,\,a}_{\pi,\,\nu}(a_1,i)\,.
\label{RPAA}
\end{equation}
Our Eq.\,(\ref{PCAC}) is in agreement with the general discussion
\cite{CS} of the structure of the matrix elements of two
currents.

In the next step, we verify the CVC equation for the hadron
part of our RMC amplitude. For this
purpose, we calculate separately the divergence of the weak
vector and axial parts of this amplitude with the
result
\begin{eqnarray}
k_\nu\,[\,M^{B,\,a}_{V,\,\mu \nu}(1)\,+\,M^{B,\,a}_{\mu \nu}(5)
\,+\,M^{B,\,a}_{\mu \nu}(6)\,]\,=\,i\varepsilon^{3\,a\,b}
\bar u(p')\,\hat J^b_{V,\,\mu}(q_1)\,u(p)\, \label{EMVCC} \\
k_\nu\,[\,M^{B,\,a}_{A,\,\mu \nu}(1)\,+\,\sum_{i=2}^4\,M^{B,\,a}_{\mu
\nu}(i)\,+\,M^a_{\mu \nu}(\pi)\,+\,M^{c.\,t.,\,a}_{\mu \nu}(a_1)]\,=\,
i\varepsilon^{3\,a\,b}
\bar u(p')\,\hat J^b_{A,\,\mu}(q_1)\,u(p)\,, \label{EMACC}
\end{eqnarray}
which leads to the correct continuity equation
\begin{equation}
k_\nu\,[M^{B,\,a}_{\mu \nu}\,+\,M^a_{\mu \nu}(\pi)\,+\,
M^{c.\,t.,\,a}_{\mu \nu}(a_1)]\,=\,
i\varepsilon^{3\,a\,b} \bar u(p')\,\hat J^b_{W,\,\mu}(q_1)\,u(p)
\,. \label{CVC}
\end{equation}

The gauge invariance of the combination $T^1\,-\,iT^2$ can be now
verified simply by changing $\varepsilon^*_\nu(k)\,\rightarrow\,
k_\nu$ in Eqs.\,(\ref{Ta}) and (\ref{Ma})
and using Eq.\,(\ref{CVC}).

%% file: chapter3.tex

\section{Discussion and the results \label{CH3}}

Equations (\ref{PCAC}) and (\ref{CVC}) are in agreement with the
general results obtained in Ref.\,\cite{CS} for the matrix
elements of two current. Let us note that our amplitude satisfy
them exactly. The derivation of such an amplitude based on the
low-energy theorems \cite{AD}-\cite{BF} provides it up to the
terms linear in $k$ and $q$.

As we have discussed in Ref.\,\cite{STG}, the form of the
Lagrangian (\ref{LNLNPNPR}) from which the RMC amplitude
is constructed, is restricted at threshold by PCAC and current
algebras. It follows that our amplitude coincides at
threshold with the one obtained from the low-energy theorems.
At higher energies (up to $1\, GeV$) it is demanded
that the HLS approach incorporates vector meson
dominance, respects the Weinberg sum rules and the KSFR relation
and describes reasonably
physical processes such as $\rho\,\rightarrow\,\pi\pi$,
$a_1\,\rightarrow\,\rho \pi$ etc. It was also shown in \cite{STG}
that only at energies $\sim\,0.8\,GeV$ one can expect sizeable
effects depending on the chosen Lagrangian model. Then in the region of
energies relevant for the process of RMC in nuclei one can
consider our transition amplitude as reliably fixed.

Let us compare our radiative hadron amplitude with the results
of Ref.\,\cite{CS}. We start with the amplitudes
$M^{B,\,a}_{\mu \nu}(1-6)$ from Eqs.\,(\ref{MBa1})-(\ref{MBa6}).
The amplitude $M^{B,\,a}_{\mu \nu}(1)$ with the nucleon electroweak
currents (\ref{JaW}) and (\ref{Jem}) contains the standard
nucleon Born amplitude and also a contribution due to the
non-pole part of the induced pseudoscalar. Actually, it comes
from the transverse part of the first term of the nucleon axial
current $\hat J^a_{A,\,\mu}$ Eq.\,(\ref{JaA}), because the term
$-ig_A \Delta^{a_1}_F(q_1)q_{1 \mu}q_{1 \eta} \gamma_\eta
\gamma_5$ has the form of the induced pseudoscalar with the form
factor $-ig_A \Delta^{a_1}_F(q_1)q_{1 \eta} \gamma_\eta$. For the
nucleon on the mass shell, this is effectively
$2M g_A \Delta^{a_1}_F(q_1)q_{1 \mu} \gamma_5$ which allows one to
write the induced pseudoscalar term with the
effective form factor
$\Delta^\pi_F(q_1)\,\rightarrow\,\Delta^\pi_F(q_1)-\Delta^{a_1}_F(q_1)$.

The large term $M^{B,\,a}_{\mu \nu}(2)$ of Eq.\,(\ref{MBa2})
cancels at threshold with the contact term $M^{c.\,t.,\,a}_{\mu
\nu}(a_1,1)$ of Eq.\,(\ref{Ma11}) and only higer order terms
survive (see below). The amplitude $M^{B,\,a}_{\mu \nu}(3)$ does
not contribute to the considered reaction. The terms
$M^{B,\,a}_{\mu \nu}(4)$ and $M^{B,\,a}_{\mu \nu}(5)$ are present
in Eq.\,(58) and Eq.\,(59), respectively.

The amplitude $M^{B,\,a}_{\mu \nu}(6)$ is due to vector $\rho \rho
\rho$ interaction  (Fig.\,1d) and it contributes in the leading
order to the terms linear in $k$ and $q$ as
\begin{equation}
\Delta\,M^{B,\,a}_{\mu \nu}(6)\, \approx\, \frac{1}{m^2_\rho}
\varepsilon^{3\,a\,b}\bar
u(p')\frac{\tau^b}{2}[\,(q_\eta+k_\eta)\gamma_\eta\,\delta_{\mu
\nu}\,+\,(k_\nu-2q_\nu)\gamma_\mu\,+\,(q_\mu-2k_\mu)\gamma_\nu\,]
u(p)\,.  \label{del1}
\end{equation}

Next we discuss the amplitudes $M^a_{\mu \nu}(\pi)$ of
Eqs.\,(\ref{Mmca})-(\ref{Mapi5}). As we have already mentioned
$M^{m.\,c.,\,a}_{\mu \nu}$ is the standard mesonic current.
Having in mind that in the considered symmetry scheme
$m^2_{a_1}\,=\,2 m^2_\rho$, we can sum up
\begin{equation}
M^a_{\mu \nu}(\pi,1)\,+\,M^a_{\mu \nu}(\pi,4)\,=\,iMg_A
\Delta^\pi_F(q_1)\,\varepsilon^{3\,a\,b}\Gamma^b(p',p)
\delta_{\mu \nu}\,.
\end{equation}
Such a term is present in Eq.\,(59) of \cite{CS}.

The amplitudes $M^a_{\mu \nu}(\pi,2)$, $M^a_{\mu \nu}(\pi,3)$
and $M^a_{\mu \nu}(\pi,5)$ contribute in higher order in k and q.
These terms cannot be obtained in the standard expansion
technique \cite{CS}.

As we have already mentioned, the large term $M^{c.\,t.,\,a}_{\mu
\nu}(a_1,1)$ compensates in the leading order $M^{B,\,a}_{\mu
\nu}(2)$, the result of the sum providing the terms
\begin{eqnarray}
\Delta(1)\, & \equiv & \,M^{B,\,a}_{\mu\nu}(2)\,+\,M^{c.\,t.,\,a}_{\mu
\nu}(a_1,1)\,\approx\,
\frac{g_A}{4m^2_\rho}\,q_\mu \Delta^\pi_F(q)[\,q_\nu q_\zeta -
q^2\delta_{\nu \zeta} \nonumber \\
& & \,+k_\nu k_\zeta + 2(k \cdot q)\delta_{\nu \zeta}
-k_\nu q_\zeta - k_\zeta q_\nu\,]\varepsilon^{3\,a\,b}
\Gamma^b_\zeta (p',p)\,.  \label{del2}
\end{eqnarray}

In the same order, the terms $M^{c.\,t.,\,a}_{\mu \nu}(a_1,2)$
and $M^{c.\,t.,\,a}_{\mu \nu}(a_1,3)$ contribute as
\begin{eqnarray}
\Delta(2) & \,\equiv & \,M^{c.\,t.,\,a}_{\mu \nu}(a_1,2)\,+\,
M^{c.\,t.,\,a}_{\mu \nu}(a_1,3)\,\approx\,
\frac{g_A}{4m^2_\rho}\,q_\mu \Delta^\pi_F(q)\{\,2[\,q_\nu k_\zeta
-(q \cdot k)\delta_{\nu \zeta}\,] \nonumber \\
& & +\,(\,q^2\delta_{\nu \zeta}- q_\nu q_\zeta\,)\,+\,[\,k_\nu
q_\zeta - (k \cdot q)\delta_{\nu \zeta}\,]\}
\varepsilon^{3\,a\,b}\Gamma^b_\zeta (p',p)\,. \label{del3}
\end{eqnarray}
We now sum up the results (\ref{del2}) and (\ref{del3}) and leave
the terms linear in $k$ and $q$ only
\begin{equation}
\Delta(1)\,+\,\Delta(2)\,=\,
\frac{g_A}{4m^2_\rho}\,q_\mu \Delta^\pi_F(q) [\,q_\nu k_\zeta
-(q \cdot k)\delta_{\nu \zeta}\,]
\varepsilon^{3\,a\,b}\Gamma^b_\zeta (p',p)\,. \label{del12}
\end{equation}
The terms of this order are present also in Eq.\,(58) of
Ref.\,\cite{CS}. Our model provides the amplitudes $\bar V_i$
consistently.

At last, the sum of the amplitudes $M^{c.\,t.,\,a}_{\mu \nu}(a_1,4)$
and $M^{c.\,t.,\,a}_{\mu \nu}(a_1,5)$  contributes in the order
\begin{equation}
\Delta(3)\,=\,
 \frac{g_A}{4 m^2_\rho}[\,(k_\mu+q_\mu)
\delta_{\nu \zeta}\,-\,(2k_\zeta+q_\zeta)\delta_{\mu \nu}\,+\,
(2q_\nu-k_\nu)\delta_{\mu \zeta}\,]\varepsilon^{3\,a\,b}
\Gamma^b_\zeta(p',p)\,. \label{del4}
\end{equation}

In conclusion we notice that
\begin{enumerate}
\item Our amplitude for RMC derived from the chiral Lagrangian of the HLS
satisfies PCAC, CVC and gauge invariance exactly and coincides in the
leading order with the standard one.
\item Our resulting correction terms
linear in $k$ and $q$ (see
Eqs.\,(\ref{del1}),(\ref{del12}) and (\ref{del4})) differ
from those obtained in \cite{AD}-\cite{KL} by the standard
expansion technique. This is due to the different
prescription for passing towards higher energies.
In our approach \cite{STG}, the vector meson dominance, Weinberg sum rules
and KSFR relation restrict the physical amplitudes at higher
energies.
\item One can obtain explicitly higher order terms from our amplitude,
which is not possible using the low energy expansion technique.
\end{enumerate}

%% file: ack.tex
\vspace{10pt}

This work is supported in part by the grant GA \v{C}R 202/97/0447.
The research of F.\,C.\,K.\, is supported in part by NSERCC.

We thank H.\,W.\, Fearing for a critical reading of the paper.

%% file: ref.tex

%% file: fig.tex
\newpage
\vskip 20pt
\hskip 20pt
\begin{picture}(10000,15000)
\drawline\fermion[\NE\REG](0,0)[6000]
\drawarrow[\NE\ATBASE](\pmidx,\pmidy)
\global\advance\fermionfrontx by 900
\put(\fermionfrontx,\fermionfronty){$p$}
\put(\fermionbackx,\fermionbacky){\circle*{800}}
\put(\fermionbackx,-1800){a}
\drawline\photon[\SE\REG](\fermionbackx,\fermionbacky)[7]
\global\advance\pmidy by -270
\drawarrow[\NW\ATBASE](\pmidx,\pmidy)
\global\advance\pmidx by 900
\put(\pmidx,\pmidy){$\hat{J}^{a}_{W,\,\mu}$}
\global\advance\photonbackx by -1500
\global\advance\photonbacky by -150
\put(\photonbackx,\photonbacky){$q$}
\drawline\fermion[\N\REG](\photonfrontx,\photonfronty)[5948]
\drawarrow[\N\ATBASE](\pmidx,\pmidy)
\global\advance\pmidx by 450
\global\advance\pmidy by -300
\put(\pmidx,\pmidy){$P$}
\drawline\photon[\NE\REG](\fermionbackx,\fermionbacky)[7]
\put(\fermionbackx,\fermionbacky){\circle*{800}}
\global\advance\pmidy by -225
\drawarrow[\NE\ATBASE](\pmidx,\pmidy)
\global\advance\pmidx by 900
\global\advance\pmidy by -300
\put(\pmidx,\pmidy){$\hat{J}^{e.\,m.\,}_{\nu}$}
\global\advance\photonbackx by -1500
\global\advance\photonbacky by -300
\put(\photonbackx,\photonbacky){$k$}
\drawline\fermion[\NW\REG](\photonfrontx,\photonfronty)[6000]
\drawarrow[\NW\ATBASE](\pmidx,\pmidy)
\global\advance\fermionbackx by 900
\global\advance\fermionbacky by -300
\put(\fermionbackx,\fermionbacky){$p^{\,\prime}$}
\end{picture}
\hskip 1.cm
\begin{picture}(10000,15000)
\drawline\fermion[\NE\REG](0,0)[6000]
\drawarrow[\NE\ATBASE](\pmidx,\pmidy)
\global\advance\fermionfrontx by 900
\put(\fermionfrontx,\fermionfronty){$p$}
\put(\fermionbackx,-1800){b}
\put(\fermionbackx,\fermionbacky){\circle*{800}}
\drawline\photon[\NE\REG](\fermionbackx,\fermionbacky)[9]
\global\advance\pmidx by 1800
\global\advance\pmidy by 1600
\drawarrow[\NE\ATBASE](\pmidx,\pmidy)
\global\advance\photonbackx by -1500
\global\advance\photonbacky by -300
\put(\photonbackx,\photonbacky){$k$}
\drawline\fermion[\N\REG](\photonfrontx,\photonfronty)[5948]
\drawarrow[\N\ATBASE](\pmidx,\pmidy)
\global\advance\pmidx by -1350
\global\advance\pmidy by -300
\put(\pmidx,\pmidy){$Q$}
\drawline\fermion[\NW\REG](\fermionbackx,\fermionbacky)[6000]
\drawarrow[\NW\ATBASE](\pmidx,\pmidy)
\put(\fermionfrontx,\fermionfronty){\circle*{800}}
\global\advance\fermionbackx by 900
\global\advance\fermionbacky by -300
\put(\fermionbackx,\fermionbacky){$p^{\,\prime}$}
\drawline\photon[\SE\REG](\fermionfrontx,\fermionfronty)[9]
\global\advance\pmidx by 1750
\global\advance\pmidy by -2100
\drawarrow[\NW\ATBASE](\pmidx,\pmidy)
\global\advance\photonbackx by -1800
\global\advance\photonbacky by -150
\put(\photonbackx,\photonbacky){$q$}
\end{picture}
\hskip 1.cm
\begin{picture}(10000,15000)
\drawline\fermion[\N\REG](0,0)[13948]
\drawarrow[\N\ATBASE](0,2000)
\drawarrow[\N\ATBASE](0,11948)
\put(0,-1800){c}
\global\advance\fermionbackx by -1500
\put(\fermionbackx,\fermionbacky){$p^{\prime}$}
\put(\fermionbackx,0){$p$}
\global\advance\pmidy by -106
\drawline\fermion[\SE\REG](\pmidx,\pmidy)[2000]
\drawline\fermion[\NE\REG](\fermionbackx,\fermionbacky)[150]
\drawline\fermion[\NW\REG](\fermionbackx,\fermionbacky)[2000]
\drawline\fermion[\W\REG](\fermionbackx,\fermionbacky)[106]
\thicklines
\drawline\fermion[\SE\REG](\fermionbackx,\fermionbacky)[2075]
\thinlines
\drawline\photon[\SE\REG](\fermionbackx,\fermionbacky)[8]
\global\advance\pmidx by 900
\global\advance\pmidy by -1200
\drawarrow[\NW\ATBASE](\pmidx,\pmidy)
\global\advance\photonbackx by -1800
\put(\photonbackx,\photonbacky){$q$}
\global\advance\photonbackx by 1200
\global\advance\photonbacky by 1500
\put(\photonbackx,\photonbacky){$\hat{J}^{\,a}_{W,\,\mu}$}
\drawline\fermion[\NE\REG](0,6974)[2000]
\drawline\fermion[\NW\REG](\fermionbackx,\fermionbacky)[150]
\drawline\fermion[\SW\REG](\fermionbackx,\fermionbacky)[1900]
\drawline\fermion[\S\REG](\fermionbackx,\fermionbacky)[106]
\put(0,6974){\circle*{800}}
\thicklines
\drawline\fermion[\NE\REG](\fermionbackx,\fermionbacky)[2075]
\thinlines
\drawline\photon[\NE\REG](\fermionbackx,\fermionbacky)[8]
\global\advance\pmidx by 1000
\global\advance\pmidy by 700
\drawarrow[\NE\ATBASE](\pmidx,\pmidy)
\global\advance\photonbackx by -1200
\put(\photonbackx,\photonbacky){$k$}
\global\advance\photonbackx by 800
\global\advance\photonbacky by -1700
\put(\photonbackx,\photonbacky){$\hat{J}^{\,e.\,m.}_\nu$}
\global\advance\photonfrontx by 600
\global\advance\photonfronty by -600
\put(\photonfrontx,\photonfronty){$B_{\,1}$}
\global\advance\photonfronty by -2200
\put(\photonfrontx,\photonfronty){$B_{\,2}$}
\end{picture}
\vskip 50pt
\hskip 80pt
\begin{picture}(10000,15000)
\drawline\fermion[\N\REG](0,0)[13948]
\drawarrow[\N\ATBASE](0,2000)
\drawarrow[\N\ATBASE](0,11948)
\put(0,-1800){d}
\global\advance\fermionbackx by -1500
\put(\fermionbackx,\fermionbacky){$p^{\prime}$}
\put(\fermionbackx,0){$p$}
\global\advance\pmidy by 75
\drawline\fermion[\E\REG](0,\pmidy)[1400]
\global\advance\pmidy by 600
\put(\pmidx,\pmidy){$\rho$}
\drawline\fermion[\N\REG](\fermionbackx,\fermionbacky)[50]
\drawline\fermion[\NE\REG](\fermionbackx,\fermionbacky)[2000]
\drawline\fermion[\SE\REG](\fermionbackx,\fermionbacky)[75]
\drawline\fermion[\SW\REG](\fermionbackx,\fermionbacky)[2000]
\drawline\fermion[\N\REG](\fermionbackx,\fermionbacky)[50]
\thicklines
\drawline\fermion[\NE\REG](\fermionfrontx,\fermionfronty)[2000]
\thinlines
\drawline\photon[\NE\REG](\fermionbackx,\fermionbacky)[8]
\global\advance\pmidx by 1000
\global\advance\pmidy by 700
\drawarrow[\NE\ATBASE](\pmidx,\pmidy)
\global\advance\photonbackx by -1200
\put(\photonbackx,\photonbacky){$k$}
\global\advance\photonbackx by 800
\global\advance\photonbacky by -1700
\put(\photonbackx,\photonbacky){$\hat{J}^{\,3}_{V,\,\nu}$}
\drawline\fermion[\S\REG](\fermionfrontx,\fermionfronty)[100]
\drawline\fermion[\NE\REG](\fermionbackx,\fermionbacky)[75]
\drawline\fermion[\SE\REG](\fermionbackx,\fermionbacky)[2000]
\drawline\fermion[\SW\REG](\fermionbackx,\fermionbacky)[100]
\drawline\fermion[\NW\REG](\fermionbackx,\fermionbacky)[2000]
\drawline\fermion[\NE\REG](\fermionbackx,\fermionbacky)[75]
\drawline\fermion[\W\REG](\fermionbackx,\fermionbacky)[1400]
\drawline\fermion[\N\REG](\fermionbackx,\fermionbacky)[40]
\thicklines
\drawline\fermion[\E\REG](\fermionbackx,\fermionbacky)[1400]
\drawline\fermion[\SE\REG](\fermionbackx,\fermionbacky)[2000]
\thinlines
\global\advance\pmidx by 400
\global\advance\pmidy by 800
\put(\pmidx,\pmidy){$\rho$}
\global\advance\pmidx by -1600
\global\advance\pmidy by -1600
\put(\pmidx,\pmidy){$\rho$}
\drawline\photon[\SE\REG](\fermionbackx,\fermionbacky)[8]
\global\advance\pmidx by 900
\global\advance\pmidy by -1200
\drawarrow[\NW\ATBASE](\pmidx,\pmidy)
\global\advance\photonbackx by -1800
\put(\photonbackx,\photonbacky){$q$}
\global\advance\photonbackx by 1200
\global\advance\photonbacky by 1500
\put(\photonbackx,\photonbacky){$\hat{J}^{\,a}_{V,\,\mu}$}
\end{picture}
\hskip 1.cm
\begin{picture}(10000,15000)
\drawline\fermion[\NE\REG](0,0)[6000]
\drawarrow[\NE\ATBASE](\pmidx,\pmidy)
\global\advance\fermionfrontx by 600
\put(\fermionfrontx,\fermionfronty){$p$}
\put(\fermionbackx,-1800){e}
\drawline\fermion[\SE\REG](\fermionbackx,\fermionbacky)[6000]
\drawarrow[\SE\ATBASE](\pmidx,\pmidy)
\global\advance\fermionbackx by -2000
\put(\fermionbackx,\fermionbacky){$p^{\prime}$}
\drawline\scalar[\N\REG](\fermionfrontx,\fermionfronty)[3]
\drawarrow[\S\ATBASE](\pmidx,\pmidy)
\global\advance\pmidx by 450
\put(\pmidx,\pmidy){$q_{1}$}
\global\advance\pmidx by -2000
\put(\pmidx,\pmidy){$B$}
\drawline\photon[\NE\FLIPPED](\scalarbackx,\scalarbacky)[7]
\global\advance\pmidx by -300
\drawarrow[\NE\ATBASE](\pmidx,\pmidy)
\global\advance\pmidx by 1200
\put(\pmidx,\pmidy){$\hat{J}^{e.\,m.\,}_{\nu}$}
\global\advance\photonbackx by -1200
\put(\photonbackx,\photonbacky){$k$}
\drawline\photon[\NW\REG](\scalarbackx,\scalarbacky)[7]
\global\advance\pmidx by -2500
\put(\pmidx,\pmidy){$\hat{J}^{a}_{A,\,\mu}$}
\global\advance\pmidx by 2600
\global\advance\pmidy by 250
\drawarrow[\SE\ATBASE](\pmidx,\pmidy)
\global\advance\photonbackx by 1200
\global\advance\photonbacky by -300
\put(\photonbackx,\photonbacky){$q$}
\put(\scalarbackx,\scalarbacky){\circle*{800}}
\end{picture}
\vskip 1.cm
\begin{center}
Fig.\,1. The radiative hadron amplitude. Graphs a,b -- the nucleon
Born terms; graphs c,d -- the contact terms related to the nucleon
Born terms, the possible pairs $(B_1,B_2)$ are $\rho,\pi$,
$\omega,\pi$ and $\rho,\rho$; graph e -- the contact terms of the
$B\,=\,\pi$ or $a_1$ range.
\end{center}